\documentclass[12pt]{article}

\usepackage[english]{babel}
\usepackage{amsmath}
\usepackage{graphicx}
\usepackage{color}
\usepackage[usenames,dvipsnames,svgnames,table]{xcolor}
\usepackage{setspace}
\makeatletter
\renewcommand{\@biblabel}[1]{\quad#1.}
\makeatother

\date{}

\begin{document}
%\maketitle

\begin{flushleft}
{\Large
\textbf{Hybrid RHF/MP2 geometry optimizations with the effective fragment molecular orbital method}
}
\\
Anders S. Christensen$^1$,
Casper Steinmann$^2$,
Dmitri G. Fedorov$^3$,
Jan H. Jensen$^1$
\\
\bf{1} Department of Chemistry, University of Copenhagen, Copenhagen, Denmark
\\
\bf{2} Department of Physics, Chemistry and Pharmacy, University of Southern Denmark, Odense, Denmark
\\
\bf{3} Nanosystem Research Institute (NRI), National Institute of Advanced Industrial Science and Technology (AIST), Tsukuba, Ibaraki, Japan
\\
$\ast$ corresponding author, E-mail: jhjensen@chem.ku.dk
\end{flushleft}

\begin{abstract}
The frozen domain effective fragment molecular orbital method is extended to allow
for the treatment of a single fragment at the MP2 level of theory. The approach is
applied to the conversion of chorismate to prephenate by chorismate mutase, where
the substrate is treated at the MP2 level of theory while the rest of the system is
treated at the RHF level.
MP2 geometry optimization is found to lower the barrier by up to 3.5 kcal/mol compared
to RHF optimzations and ONIOM energy refinement and leads to a smoother convergence
with respect to the basis set for the reaction profile. For double zeta basis sets the
increase in CPU time relative to RHF is roughly a factor of two.
\end{abstract}

\section{Introduction}
Combined quantum mechanical/molecular mechanical (QM/MM) and fragment-based quantum mechanical methods\cite{MCF,XPol,ELG1,MFCC,SMFEFP,EFP,DCrev,KEM,PMISP,MTA,ADMA,EEMB,fragrev}, recently reviewed \cite{fragrev,qmmrev}, are becoming increasingly popular for large molecular systems. 
In the fragment molecular orbital method (FMO) \cite{FMO01,FMOrev1,FMOrev2} one does fragment calculations in the presence of the embedding potential of all the other fragments, whereas in the recently developed effective fragment molecular orbital method (EFMO)\cite{efmo1,efmo2} fragment polarizabilities are used instead to approximate the many-body polarization.

For fast geometry optimizations, FMO with the frozen domain and dimers (FDD) \cite{fd-fdd} has been proposed and EFMO/FDD has been used to map the reaction path of the conversion of chorismate to prephanate in Chorismate Mutase at the RHF level for geometry optimization combined with ONIOM for energy refinement.\cite{efmom} Chorismate Mutase has also been studied extensively by many groups. Particularly, the group of Mulholland has invested considerable amount of resources to evaluate the catalytic effect of Chorismate Mutase.\cite{Mulholland1,Mulholland2,Mulholland3,Mulholland4,Mulholland5,Mulholland6,Mulholland7}
Other related QM/MM work on Chorismate Mutase includes FMO energetics refinement by Ishida et al. \cite{FMOCl} and the work of Claeyssens~{\em et al.}\cite{Thiel} who used linear scaling coupled cluster methods to obtain the reaction barrier on structures optimized using a QM/MM approach with density functional theory used to describe the QM region. 
Our recent study\cite{efmom} emphasizes that in addition to a high quality energy evaluation on the reaction complex, a conformational sampling of the reaction complex geometry is needed in order to obtain a reliable energy barrier, since the reaction barrier can fluctuate by up to 15 kcal/mol between geometry optimizations on different starting conformations.

Our previous method was to estimate the reaction barrier in Chorismate Mutase using an EFMO-RHF geometry optimization with an ONIOM MP2 energy correction\cite{efmom}.
It was clear, however, that the RHF based optimization did not always lead to a reliable MP2 correction.
In this work, we extend EFMO/FDD to enable treatment of only one fragment at the MP2 level and show that it is a good compromise between efficiency and accuracy.
Note that the effects of conformational sampling are not investigated in this paper.

This paper is organized as follows: First we present the EFMO method and our extension to the EFMO energy and gradient.
Second we compare our method to similar ONIOM calclations on the reaction barrier of the conversion of chorismate to prephanate in Chorismate Mutase.

\section{Theory}
The basics of EFMO can be summarized as follows. The system is divided into fragments and
we use the adaptive frozen orbital technique (AFO) \cite{AFO} to treat fragment boundaries
by freezing the molecular orbitals corresponding to detached covalent bonds.
\textit{Ab initio} calculations of fragments are carried out without embedding, and the 
total polarization is evaluated using fragment polarizabilities. 
In the next step, \textit{ab initio} calculations of dimers are carried out to account for
two-body quantum effects such as the charge transfer between fragment pairs within a cut-off distance, $R_\mathrm{resdim}$. The total energy in the two-body EFMO expansion is then:
\begin{eqnarray}
    E^\mathrm{EFMO} &=& \sum_I E^0_I
    + \sum_{I>J}^{R_{IJ}\leq R_\mathrm{resdim}} \left(\Delta E_{IJ}^0 - E_{IJ}^\mathrm{POL} \right)
    + \sum_{I>J}^{R_{IJ} > R_\mathrm{resdim}} E^\mathrm{ES}
    + E^\mathrm{POL}_\mathrm{tot}.
\end{eqnarray}

Here  $E^0_I$ is the quantum mechanical gas-phase energy of each monomer fragment, $\Delta E_{IJ}^0$ is the quantum mechanical two-body polarization energy between two fragments,  $E_{IJ}^\mathrm{POL}$ is the classical two-body polarization energy between two fragments, and $E^\mathrm{POL}_\mathrm{tot}$ is the classical polarization energy of the system.

In the frozen domain method,\cite{fd-fdd} the geometry of the molecular system is optimized only for a smaller subsystem called the active domain, while the atoms in the rest of the system are fixed.\\
For a given molecular system, we define two domains $F$ ("frozen") and $A$ ("active"). Domain $F$ is defined as all atoms having a frozen geometry and domain $A$ is defined as all atoms whose positions are optimized. 
Each domain is further divided into a number of molecular fragments and the EFMO energy is given by\cite{efmom}

\begin{eqnarray}
	E^{\mathrm{EFMO}} &=& %E^0_\mathrm{F}
    				    E^0_\mathrm{b}
                        + E^0_\mathrm{A}
                       %+ E^0_\mathrm{F/B}
                        + E^0_\mathrm{F/A}
                        + E^0_\mathrm{A/b}
                        + E^\mathrm{POL}_\mathrm{tot},
\end{eqnarray}
where $E^0_\mathrm{b}$ and $E^0_\mathrm{A}$ are the internal energies of domains $b$ and $A$, respectively, $E^0_\mathrm{F/A}$ is the interaction between domains $F$ and $A$,  $E^0_\mathrm{A/b}$ is the interaction between domains $A$ and $b$ and $E^\mathrm{POL}_\mathrm{tot}$ is the classical total polarization energy of the whole system. In our EFMO-RHF:MP2 extension, we evaluate the internal energies of domain $b$ and $A$ at the RHF level.
Furthermore, we specify a single fragment $H$ ("high level") from the active domain to be treated at the MP2 level of theory (see Fig. 1 for a schematic overview).
The total EFMO-RHF:MP2 energy is then given as

\begin{eqnarray}
	E^{\mathrm{EFMO-RHF:MP2}} &=& %E^\mathrm{0,RHF}_\mathrm{F}
                                 E^\mathrm{0,RHF}_\mathrm{b}
                                + E^\mathrm{0,RHF}_\mathrm{A}
                               %+ E^0_\mathrm{F/B}
                                + E^\mathrm{0,RHF}_\mathrm{F/A}
                                + E^\mathrm{0,RHF}_\mathrm{A/b}
                                + E^\mathrm{POL}_\mathrm{tot}
                                + E^\mathrm{0,MP2}_\mathrm{H \in A},
\end{eqnarray}
where $E^\mathrm{0,MP2}_\mathrm{H \in A}$ is the MP2 correlation energy of fragment $H$.\\\\

The corresponding EFMO energy gradients of each domain in the FDD approximation:
\begin{eqnarray}
\frac{\partial E^{\mathrm{EFMO}}}{\partial x_\mathrm{A}} & = & \frac{\partial E_{\mathrm{A}  }^0}{\partial x_{\mathrm{A}}} 
									%+ \frac{\partial E_{\mathrm{H \in A}  }^\mathrm{MP2}}{\partial x_{\mathrm{A}}} 
								    + \frac{\partial E_{\mathrm{A/b}}^0}{\partial x_{\mathrm{A}}} 
                                    + \frac{\partial E_{\mathrm{F/A}}^0}{\partial x_{\mathrm{A}}} 
                                    + \frac{\partial E^{\mathrm{POL}}_{\mathrm{tot}}}{\partial x_{\mathrm{A}}} \\
 \frac{\partial E^{\mathrm{EFMO}}}{\partial x_\mathrm{b}} & = & 0\\
\frac{\partial E^{\mathrm{EFMO}}}{\partial x_\mathrm{F}} & = & 0
\end{eqnarray}
This gives the following EFMO-RHF:MP2 energy gradients:
\begin{eqnarray}
\frac{\partial E^{\mathrm{EFMO\mbox{-}RHF:MP2}}}{\partial x_\mathrm{A}} & = & 
									  \frac{\partial E^{\mathrm{EFMO}}}{\partial x_\mathrm{A}}
									+ \frac{\partial E_{\mathrm{H \in A}  }^\mathrm{MP2}}{\partial x_{\mathrm{A}}} \\
% \frac{\partial E^{\mathrm{EFMO}}}{\partial x_\mathrm{B}} & = & 0\\
\frac{\partial E^{\mathrm{EFMO\mbox{-}RHF:MP2}}}{\partial x_\mathrm{b}} & = & \frac{\partial E^{\mathrm{EFMO}}}{\partial x_\mathrm{b}} = 0\\
\frac{\partial E^{\mathrm{EFMO\mbox{-}RHF:MP2}}}{\partial x_\mathrm{F}} & = & \frac{\partial E^{\mathrm{EFMO}}}{\partial x_\mathrm{F}} = 0
\end{eqnarray}
Where $\frac{\partial E_{\mathrm{H \in A}  }^\mathrm{MP2}}{\partial x_{\mathrm{A}}}$ contains the gradient of the MP2 correlation energy for fragment H$\in$A. %Note that the gradient for atoms in b is zero only for those atoms that are not also in A.
	
\section{Methods}
All calculations were carried out in a development version of GAMESS \cite{gamess} where FMO
and EFMO are implemented \cite{FMO3}.

Starting structures for Chorismate Mutase were obtained from Steinmann \textit{et al}.\cite{efmom} who prepared the structures following Claeyssens \textit{et al}.\cite{Mulholland7}. The preparation can be summarized as follows: The experimental structure of Chorismate Mutase was obtained from the Protein Data Bank (PDB code: 2CHT) and protonated using PDB2PQR at pH 7. The inhibitors were manually replaced with Chorismate in the reactant state. The complexes were simulated in GROMACS with the CHARMM27 force field at 300K. The structure was then prepared for fragment based calculations in FragIt. \cite{fragit}
All residues with an atom within a distance of 2.0 \AA~from any atom in chorismate  were assigned to the A (active) domain. All atoms in the prephanate/chorismate reaction complex were assigned to the H fragment. See Fig. 1. The total system consists of 313 fragments, divided as 213 fragments in the frozen $F$ domain, 92 fragment in the polarizable $b$ domain, and 8 fragments in active $A$ domain of which one fragment (the reaction complex) is treated at a higher level, i.e. in the $H$ domain.

The adiabatic mapping was carried out using the presented EFMO-RHF:MP2 gradient with 6-31G(d) basis set on all atoms. Two additional runs were also carried out, in these cases with the cc-pVDZ or cc-pVTZ on chorismate and 6-31G(d) on remaining atoms.
The EFMO-RHF/6-31G(d):MP2//cc-pVTZ reaction path was obtained starting from the converged structures in the EFMO-RHF/6-31G(d):MP2//cc-pVDZ reaction path.

The RESDIM keyword was set to 1.5 and the optimization convergence criterion was set to $5.0 \cdot 10^{-4}$ Hartree/Bohr. Each step of the reaction path was obtained by imposing harmonic constraints on $R_{12}$ and $R_{13}$ with a force constant of 500 kcal/\AA.
The FDD approximation was enabled by setting MODFD=3 in all calculations.

Timings for the optimization procedure were carried out on 80 Intel Xeon X5550 CPU cores distributed across 10 nodes and the Generalized Distributed Data Interface (GDDI) was used to run the code in parallel.\cite{GDDI}

\section{Results}

\textbf{Transition State Structure.} We define the reaction coordinate similarly to Claeyssens \textit{et al.}\cite{Mulholland7}~as the difference in bond length between the breaking O2-C1 bond and the forming C4-C3 bond in chorismate, i.e. $R = R_{21} - R_{43}$ (see Fig. 2). The reaction coordinate of the transition state was found to be -0.17 \AA~using the 6-31G(d) basis set on the MP2 fragment and -0.43 \AA~for both the cc-pVTZ and cc-pVDZ basis set reaction paths. This convergence with respect to basis set is in good, quantitative agreement with the coordinates obtained by Szefczyk \textit{et al.}\cite{szefczyk}.
In comparison, the corresponding MP2:RHF ONIOM calculations by Steinmann \textit{et al.}\cite{efmom}~resulted in transition state reaction coordinates of 0.13, -0.36, and 0.13 \AA~with the cc-pVDZ, cc-pVTZ and cc-pVQZ basis sets used in the MP2 calculation, respectively  
 %As before, in contrast to the ONIOM approach, we find that for increasing basis set sizes, the transition state reaction coordinate value is systematically reduced {\color{green} reduced? with respect to what?}.

\textbf{Reaction Barrier.} Electronic energy barriers and reaction coordinates for the transition state are given in Table 1 and Fig.~3.  We find the electronic energy barrier at the EFMO-RHF/6-31G(d):MP2/6-31G(d) level of theory to be 20.95 kcal/mol. Increasing the size of the basis set on the MP2 fragment decreases the barrier to 19.21 kcal/mol with the cc-pVDZ basis set and 18.34 kcal/mol with the cc-pVTZ basis set. 

In comparison, the corresponding MP2:RHF ONIOM calculations by Steinmann \textit{et al.}~resulted in barriers of 22.24, 19.75, and 21.79 kcal/mol, respectively. In contrast to the ONIOM approach, we find that for increasing basis set sizes, the electronic energy barrier is systematically reduced.
The experimental enthalpy barrier has been measured to be 12.7 kcal/mol.\cite{kast,Mulholland7}.

\textbf{Reaction Energy.} The energy difference between the product and reactant state is found to be -3.2 kcal/mol using the 6-31G(d) basis set on chorismate. Increasing the basis set to cc-pVDZ and cc-pVTZ on chorismate decreased the reaction energy to -6.83 kcal/mol and -6.17 kcal/mol, respectively. The ONIOM approach by Steinmann \textit{et al.} found the reaction energy to be between -5.48 kcal/mol to -0.82 kcal/mol. However, in the ONIOM approach increasing the basis set from cc-pVTZ on chorismate increased the reaction energy from -5.48 kcal/mol to -1.17 kcal/mol. We find that all three basis sets are in close agreement, and only a 0.7 kcal/mol difference between the cc-pVDZ and cc-pVTZ reaction paths.

\textbf{Timings} Running on 80 cores distributed on 10 compute nodes and using the default compute node load balancing scheme, the average time for a geometry optimization step was 760s at the EFMO-RHF/6-31G(d) level of theory\cite{efmom}.
For the EFMO-RHF/6-31G(d):MP2/6-31G(d) calculation, this time increased to 1526 s per step.
Increasing the basis set on the MP2 part of the system to cc-pVDZ and cc-pVTZ increased the time to 1967 s and 18845 s, respectively (see Table 2).
The large increase in calculation time from cc-pVDZ to cc-pVTZ was found to be due to sub-optimal load balancing in GDDI during the MP2 part of the calculation. Subsequently, one optimization using the cc-pVTZ was carried out, in which the calculation of the MP2 fragment energy and gradient was distributed across all 10 nodes. This reduced the average gradient step time from 18845 s to 10911 s. In other words, the slower calculation
used 10 GDDI groups in the second (MP2) layer, whereas the faster one had 1 group,
during the monomer step. The latter run is more efficient because the MP2 fragment
was calculated by all 10 nodes, whereas in the former only by 1 node.

\section{Conclusion}
We have implemented an scheme for optimizing a reaction complex using a correlated method in the EFMO/FDD approximation.\cite{efmom}
Our method is computationally efficient when a moderately sized basis sets is used on the correlated fragment.
While our EFMO-RHF:MP2 approach does not achieve chemical accuracy in predicting enthalpy barrier of the conversion of chorismate to prephanate in chorismate mutase, we have demonstrated that our method serves as a rigorous and viable alternative to the widely used ONIOM approach.

\section*{Acknowledgements}
ASC is funded by the Novo Nordisk STAR Program. This work was in part supported by the Next Generation Super Computing Project, Nanoscience Program (NEXT, Japan) and Computational Materials Science Initiative (CMSI, Japan).

\begin{figure}
	\includegraphics[width=8.3cm]{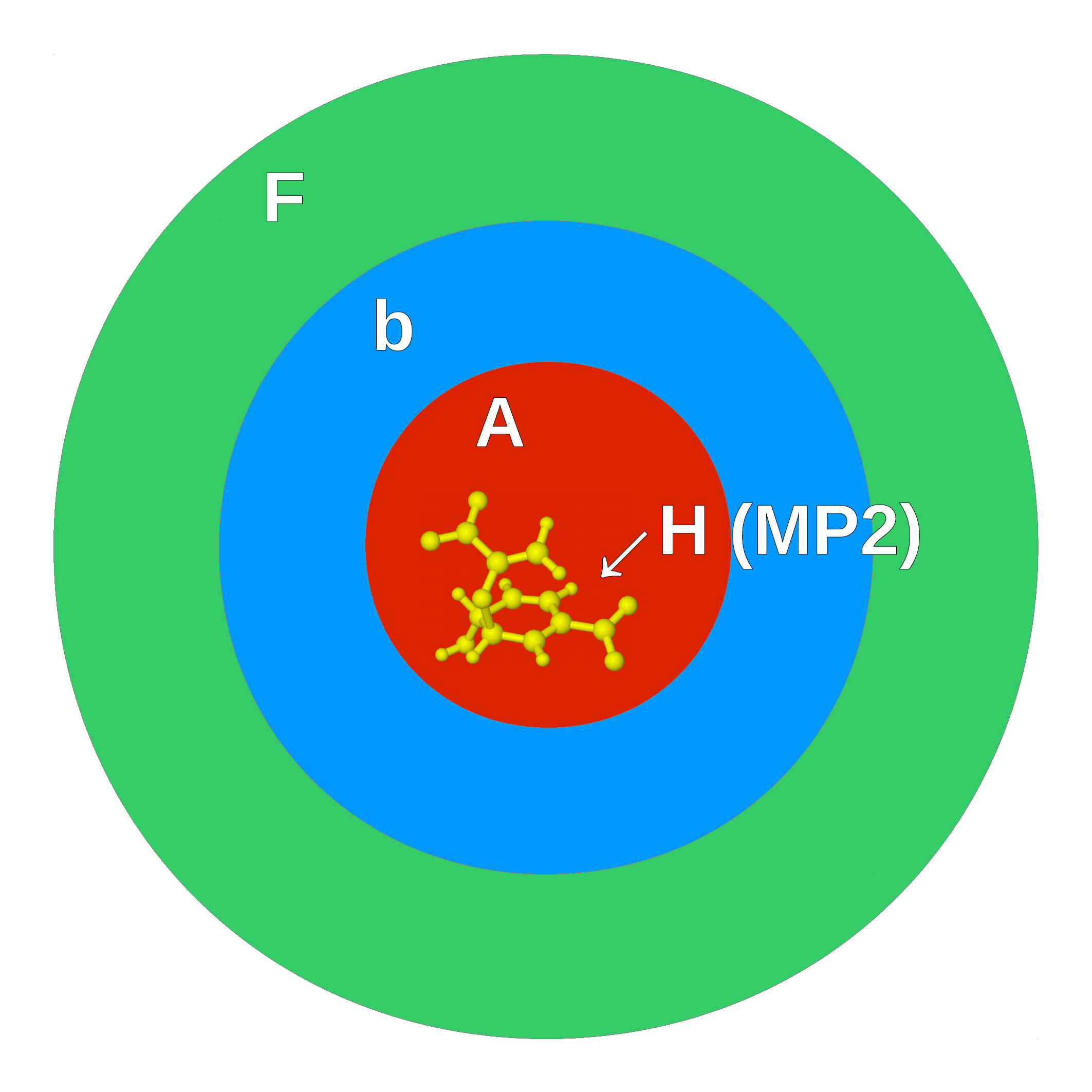}
	\label{fig:layers}
    \caption{$F$ denotes the frozen domain (green);
             $b$ denotes the polarizable domain (blue);
             $A$ denotes the active domain (red);
             $H\in A$ denotes fragment $H$, for which the 
             MP2 energy and gradients are evaluated (yellow).}
\end{figure}

\begin{figure}
	\begin{center}
    	\label{fig:corismate_reaction}
		\includegraphics[width=17.35cm]{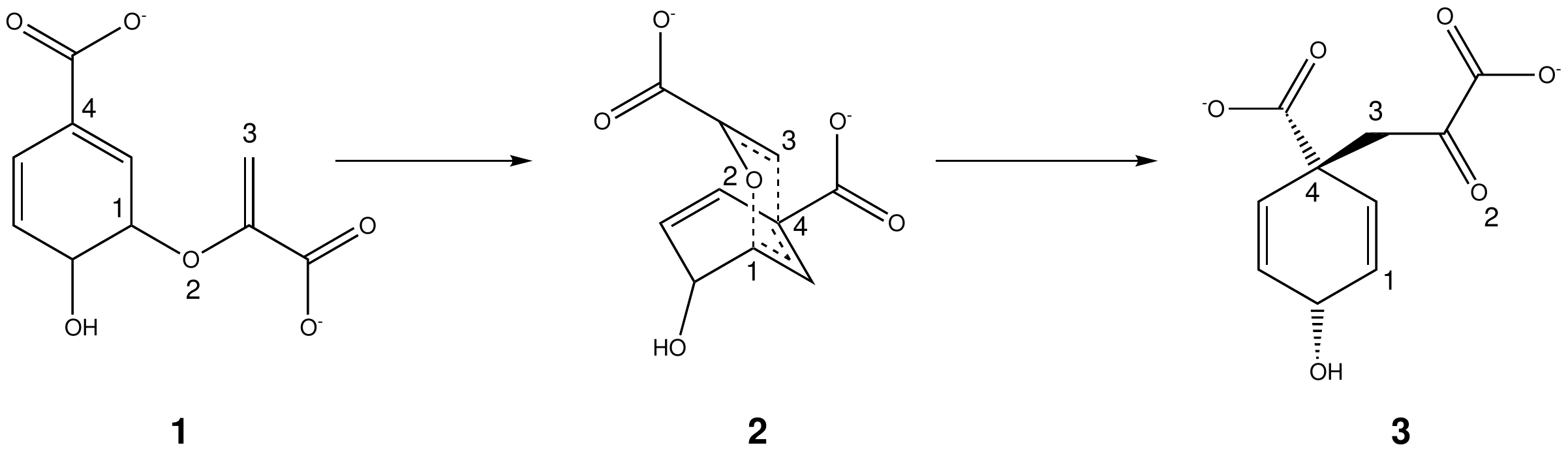}
    	\caption{Claisen rearrangement of chorismate to prephenate. 
        		 The atoms describing the reaction coordinate are 
                 marked with numbers one trough four.\cite{efmom}}
	\end{center}
\end{figure}

\begin{figure}
	\begin{center}
    	\label{fig:reaction_path}
        \includegraphics[width=8.3cm]{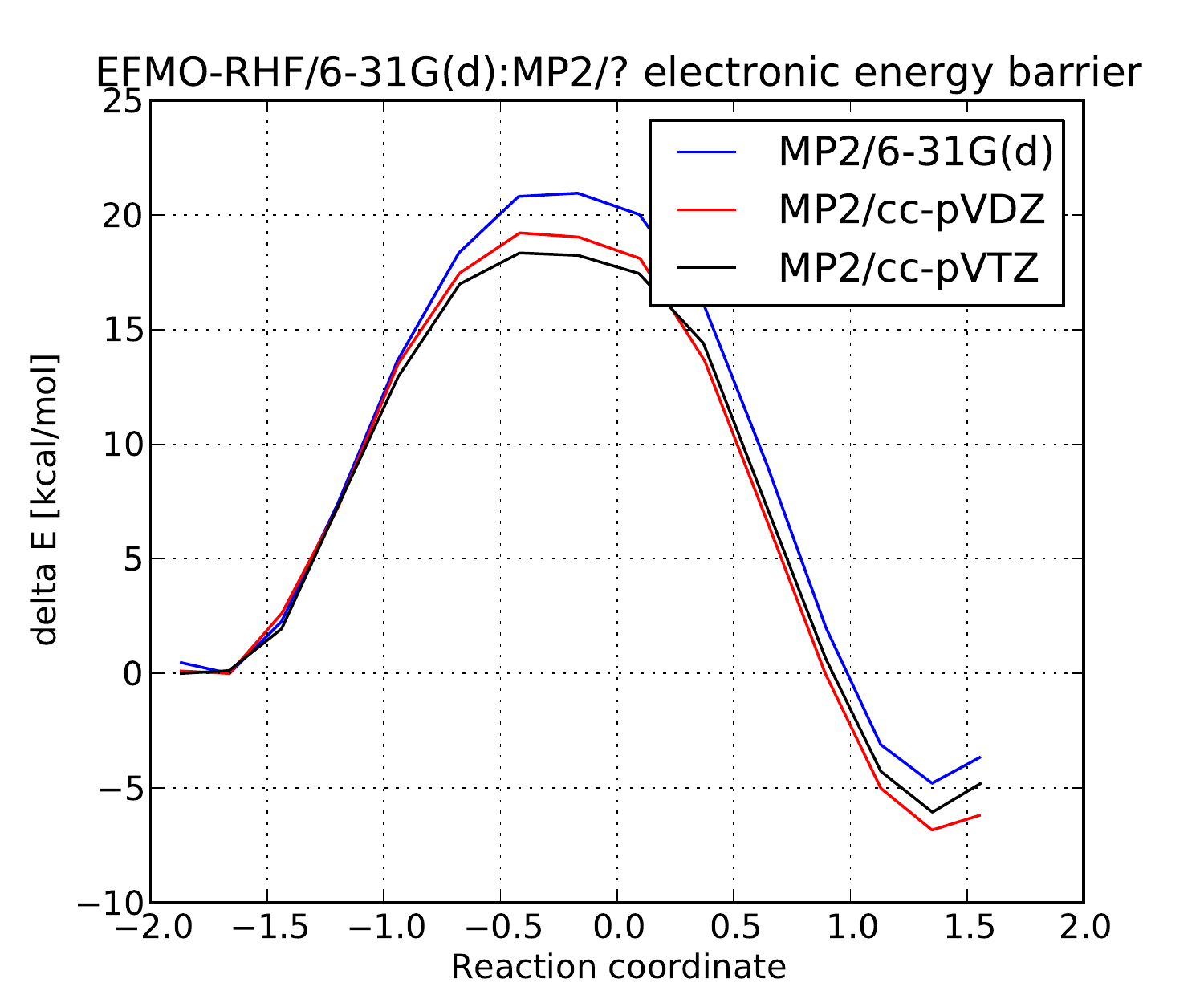}
    	\caption{Electronic energy versus reaction coordinate for the convesion of chorismate to prephanate in chorismate mutase. 
                 The three reation paths are calculated using the FDD/EFMO-RHF:MP2 approach with three different basis sets on the 
                 reaction complex in the MP2 layer. The 6-31G(d) basis set was used for the RHF layer in all three cases.}
	\end{center}
\end{figure}

\begin{table}
%\begin{center}
	\label{table:timings}
	\caption{Timings for the average geometry optimization step for Chorismate 
    		 mutase using using different methods. EFMO-RHF/6-31G(d) timings are obtained
             from Steinmann \textit{et al}.\cite{efmom} The timing marked (1 group) denotes
             that in this calculation, the MP2 part was distributed across all nodes (see text).}
	\begin{tabular}{ l r }
		Method								    	& Average step time \\\hline 
		EFMO-RHF/6-31G(d):MP2/6-31G(d) 	    		&  1527 s 			\\
		EFMO-RHF/6-31G(d):MP2/cc-pVDZ    			&  1967 s 			\\
		EFMO-RHF/6-31G(d):MP2/cc-pVTZ		 	    & 18845 s			\\
		EFMO-RHF/6-31G(d):MP2/cc-pVTZ (1 group)	& 10911 s			\\
        EFMO-RHF/6-31G(d)\cite{efmom}			    &   760 s			\\
 	\end{tabular}
%\end{center}
\end{table}

\begin{table}
\begin{center}
\label{table:energies}
    \caption{Electronic energy barrier for the conversion of prephanate to chorismate in Chorismate Mutase and the corresponding 
        	 reaction coordinate for the transition state. "EFMO" results are from the 
             presented work, calculated at the EFMO-RHF:MP2 level of theory with basis set denoted in the MP2 basis column
             for the reation complex and 6-31G(d) basis set for the rest of the system.
        "ONIOM" results are obtained from Steinmann \textit{et al}.\cite{efmom} where the structure is optimized at the
             RHF level with the 6-31G(d) basis set and MP2 with the basis set denoted in the MP2 basis column in an ONIOM correction.}
	%\begin{tabular}{ l r r r}
	%	Method											& $R$(TS)           & Energy barrier     & Reaction energy\\\hline
	%	EFMO-RHF/6-31G(d):MP2/6-31G(d)		 			& -0.17 \AA 		& 20.95 kcal/mol	 & -4.79 kcal/mol\\
	%	EFMO-RHF/6-31G(d):MP2/cc-pVDZ 					& -0.43 \AA 		& 19.21 kcal/mol 	 & -6.83 kcal/mol\\
	%	EFMO-RHF/6-31G(d):MP2/cc-pVTZ 					& -0.43 \AA 		& 18.34 kcal/mol  	 & -6.17 kcal/mol\\
    %    ONIOM(MP2/6-31G(d):RHF/6-31G(d))\cite{efmom}	&  0.13 \AA 		& 22.24 kcal/mol  	 & -3.20 kcal/mol\\
	%	ONIOM(MP2/cc-pVDZ:RHF/6-31G(d))\cite{efmom}  	& -0.36 \AA 		& 19.75 kcal/mol  	 & -5.48 kcal/mol\\
	%	ONIOM(MP2/cc-pVTZ:RHF/6-31G(d))\cite{efmom}  	&  0.13 \AA 		& 21.79 kcal/mol  	 & -1.17 kcal/mol\\
    %    ONIOM(MP2/cc-pVQZ:RHF/6-31G(d))\cite{efmom}    &  0.13 \AA 		& 21.68 kcal/mol  	 & -0.82 kcal/mol 
	%	%ONIOM(B3LYP/6-31G(d):RHF/6-31G(d)) 			&  0.13 \AA		    & 25.19 kcal/mol  	 \\
	%	%ONIOM(B3LYP/cc-pVDZ:RHF/6-31G(d))				&  0.13 \AA 		& 23.81 kcal/mol  	 \\
	%	%ONIOM(B3LYP/cc-pVTZ:RHF/6-31G(d)) 				&  0.13 \AA 		& 24.62 kcal/mol  	 \\
	%	%ONIOM(B3LYP/cc-pVQZ:RHF/6-31G(d)) 				&  0.13 \AA 		& 24.66 kcal/mol  	 \\
    %\end{tabular}
	\begin{tabular}{ l l r r r}
        Method  & MP2 basis	& $R$(TS)   & Energy barrier & Reaction energy\\
                &           &           & [kcal/mol]     & [kcal/mol]\\\hline
        EFMO    & 6-31G(d)	& -0.17 \AA & 20.95	         & -4.79\\
		EFMO    & cc-pVDZ 	& -0.43 \AA & 19.21 	     & -6.83\\
		EFMO    & cc-pVTZ 	& -0.43 \AA & 18.34  	     & -6.17\\
        ONIOM   & 6-31G(d)	&  0.13 \AA & 22.24  	     & -3.20\\
		ONIOM   & cc-pVDZ 	& -0.36 \AA & 19.75  	     & -5.48\\
		ONIOM   & cc-pVTZ 	&  0.13 \AA & 21.79  	     & -1.17\\
        ONIOM   & cc-pVQZ   &  0.13 \AA & 21.68  	     & -0.82
    \end{tabular}
\end{center}
\end{table}

\end{document}